\documentclass{llncs}
\usepackage[utf8]{inputenc}
\usepackage{graphicx}
\usepackage{todonotes}
\usepackage{amsfonts}

\title{Snitches Get Stitches:\\On The Difficulty of Whistleblowing}
\author{Mansoor Ahmed-Rengers, Ross Anderson, Darija Halatova, Ilia Shumailov}

\institute{Computer Laboratory, University of Cambridge}

\begin{document}

\maketitle
\begin{abstract}
One of the most critical security protocol problems for humans is when you are betraying a trust, perhaps for some higher purpose, and the world can turn against you if you're caught. In this short paper, we report on efforts to enable whistleblowers to leak sensitive documents to journalists more safely. Following a survey of cases where whistleblowers were discovered due to operational or technological issues, we propose a game-theoretic model capturing the power dynamics involved in whistleblowing. We find that the whistleblower is often at the mercy of motivations and abilities of others. We identify specific areas where technology may be used to mitigate the whistleblower's risk. However we warn against technical solutionism: the main constraints are often institutional.
\end{abstract}

\section{Introduction}
Whistleblowing is the act of exposing information relating to activities within an organization that are unethical, illegal or ``wrong''. Many believe that it is necessary for a healthy society that there be means by which a whistleblower can expose wrongdoing. The whistleblowing sections of government department websites reveal an interesting dissonance: they acknowledge the necessity of leaking~\cite{ukgov,boe,nidirect}, and some go as far as to offer rewards to whistleblowers~\cite{fca}. Many organizations have dedicated departments for receiving complaints, as well as policies for whistleblower protection; this extends to private firms as well.

However, time and again such internal protection mechanisms have been found lacking. A systemic failure is unsurprising, because managers do not want their failings exposed, or even to know uncomfortable facts.

Thus, as a practical matter, whistleblowers may have to leak information to external agencies such as industry regulators, the police or news organizations if they want to force reform. Telling recent examples include the Chinese medics who tried to warn the world of the COVID-19 epidemic while the political leadership was still in denial, and the many women challenging sexual harrassment in workplaces from Hollywood to Silicon Valley. A more controversial change maker was Edward Snowden. After years in which internal complaints about unlawful surveillance fell on deaf ears within the NSA and even resulted in FBI action against complainants such as Bill Binney, Ed Snowden released classified information to the world media to prove that senior officials had lied to Congress about the legality of the NSA's operations. This resulted in President Obama setting up the NSA review group, leading to reforms in how the US intelligence community operates. 

Following his disclosures, many media organisations set up supposedly secure means of making contact, using mechanisms such as PGP, Signal and Secure Drop. But do they actually work? We reviewed the advice given by a number of newspapers to potential users of their private contact mechanisms and found them to be lacking -- indeed hazardous. We set out to find a better technological solution, but soon realised we were missing the forest for the trees. Technology plays a small part in a much larger game. A review of case studies taught us that whistleblowing is really about power dynamics. In this paper, we explain what these dynamics look like and present a game-theoretic model of the relevant power relations. This leads us to present some ideas for future research.

\section{Case Studies}
\label{sec:case}
Although we looked at a wide range of case studies, we'd like to highlight a selection here to succinctly illustrate a range of scenarios. In the following section we will analyse them and see how they fit into a security game. Then we estimate weights for the different power relations and arrive at an equilibrium for each of the case studies.

\subsection{Soft and Hard Power}
First, a note about the taxonomy of power. Power is commonly seen as either soft power and hard power. Hard power is the ability to coerce, and can be based on one or more of a variety of tools: legal, financial, military, even physical intimidation. Soft power is the ability to co-opt and persuade, and its mechanisms are largely social including the credibility of a person, institution or country, public support for ideas more generally, prejudices such as sexism and racism, and even cults of personality~\cite{Nye1990}. We will use these terms as we discuss power dynamics in the following sections.

\subsection{Edward Snowden}
Edward Snowden was a contractor working for the NSA who blew the whistle on the NSA's mass data collection programs after Director of National Intelligence James Clapper lied to Congress about unlawful intelligence collection against US citizens. As an NSA (and former CIA) insider, Snowden understood the technolgies that would be needed to communicate securely with a journalist at a distance. He first got in touch with the journalist Glenn Greenwald, but Greenwald found encrypted email too annoying to use. This led Snowden to contact another journalist, Laura Poitras, who got Greenwald on board.

The Snowden case is interesting because it pits the whistleblower against one of the strongest adversaries, and because Snowden's knowledge of surveillance enabled him to disclose his identity on his own terms. But interesting as this case is, it represents a tiny minority of whistleblowing situations. Usually the adversary is not the NSA but a medium-sized business or a public-sector body such as a hospital; the whistleblower is not an expert in anonymity; and, crucially, the anonymity set (the number of people who could feasibly be the whistleblower) is not the tens of thousands of NSA technical staff, but a handful of people who knew what was going on.

\subsection{The PCAW case studies}
To get a better sense of what a ``typical'' whistleblowing event looks like we went through the case studies published by Public Concern At Work (PCAW)~\cite{pcaw}, a UK-based whistleblowing charity. Their website lists a large number of cases where employees exposed wrongdoing by their employers. In most such cases the anonymity set -- the set of people who knew the information -- was small; sometimes it was just one person. In such cases, anonymity technologies aren't going to be much help.

PCAW's advice in these cases generally revolves around finding the right authorities to talk to. Rarely do they recommend broadcasting the information to the public at large or anonymously leaking documents to a journalist. Where there are competent authorities that can keep the wrongdoer in check, this approach can make sense. A bank that's ripping off its customers can be reported to the regulator, and junior medics could deal with an incompetent surgeon in the same way. However, this often doesn't work out in practice. In the case of UK banks, the mis-selling of payment protection insurance grew into a major abuse before regulators stepped in, leading to compensation in the multiple billions; and one incompetent breast surgeon subjected more than 1,000 patients to unnecessary and damaging operations over 14 years before he was stopped~\cite{CT2020}.

PCAW's advice ignores the \emph{soft power} that the adversary may have or be able to enlist. The whistleblower may face social ostracism, difficulties in continuing work in a sector or intimidation by colluding actors. These repercussions are hard to predict and can have a severe impact. And the adverse reaction does not have to come from line managers; low-level employees can also pick on whistleblowers if they believe they have management's tacit support. In one employment tribunal case, a female employee of a Scottish government department was found gagged and tied up after complaining about sexism~\cite{scottishwhistleblower}. It can also be industry-wide. For example, the UK construction industry kept a secret blacklist of over 3,000 `undesirables' including union activists, whistleblowers and people who had raised health and safety concerns; over 40 firms paid \pounds 3,000 a year each to subscribe to the service. It was raided by the UK authorities and its operator prosecuted. It later turned out to have the covert support of the Security Service~\cite{SC2015}.

\subsection{Harvey Weinstein}
Another case where the effects of soft power are apparent is that of Harvey Weinstein, a Hollywood director who gained notoriety in 2017 when dozens of women accused him of pedatory sexual behaviour, indecent assault and rape~\cite{weinsteinNYT}. These accusations kicked off the \#MeToo movement in which a number of other rich or powerful men were publicly accused of sexual misconduct.

For years, there had been rumours about Weinstein's inappropriate behaviour, with some celebrities advising women in Hollywood to not go to his private parties~\cite{weinsteincourtney}. But Weinstein's victims rarely spoke out. This involved a mixture of hard power (fear exclusion from work in Hollywood, fear of legal costs) and soft power (the social stigma attaching to survivors of sexual assault).  In 2017 the New York Times broke the story that he'd paid off at least eight women after some thirty years of allegations of sexual harassment~\cite{KT2017}. This opened the floodgates. Dozens more women felt legally and socially safe enough to come out and make allegations of rape or indecent assault against Weinstein; and hundreds more came forward to make allegations against a range of public figures including Prince Andrew and President Donald Trump. The key to  whistleblowing on misogyny and sexual abuse in the workplace was not a new encryption technology. It was a shift in soft power that gave victims and witnesses the confidence to tell their stories in the knowledge that they had some chance of being taken seriously rather than being crushed or just ignored.

\section{Related Work}\label{sec:rwork}
There is no shortage of whistleblowing laws, organizational guidelines, and internal complaints procedures. Most of them appear aimed at damage limitation. The formal complaints procedure at the typical company will lead to the human resources department, which exists to protect the company. The usual outcome, as in Weinstein's career up to October 2017, is that complainants are intimidated into leaving and perhaps paid off.

We leave consideration of such laws and procedures to one side for now, until we have a better understanding of the power dynamics and the possible solutions. For now we focus on whistleblowing to news agencies by individuals who thereby place themselves at risk of being sued or prosecuted.



\subsection{Academic work}
Academic work on whistleblowing in the security community has focused on the usability of encryption software products~\cite{whitten1999johnny}, the design of secure messaging systems~\cite{mixminion} and so on. These tools can significantly cut the risk of leaking documents where the leaker is in a large anonymity set and the opponent is technically capable. However our focus is on how to minimise the overall risk rather on the specifics of the tools used.

\subsection{Journalistic resources}
Dozens of news organisations have web pages with guides on how to leak documents to them; see for example the New York Times \cite{nytimes_2016} and WikiLeaks \cite{wikileaks_tips}. However, they focus more on describing what makes a particular tip good and listing the acceptable communications channels, rather than on how to avoid catastrophic mistakes. For example, such pages do not provide adequate explanations of how information can leak through side channels, how to transfer physical evidence safely into digital formats, or more generally the capabilities of potential adversaries. None of them appears to help potential whistleblowers figure out what their anonymity set is and how they might expand that. 

To illustrate these shortcomings, consider the guidelines published by the New York Times. Their web page has links to Instructions and Security, and suggests four channels: WhatsApp, Signal, physical post and email.

The web page mentions that WhatsApp keeps `records of the phone numbers involved in the exchange and the users’ metadata, including timestamps on messages', but does not explain what metadata is, or how it can allow the FBI to track the leaker. For Signal, the web page mentions that Signal saves only the phone number and the time of last activity and states that no metadata about the communication gets saved. This web page also fails to explain metadata and how Signal has been compromised in the past. The new user is just not told that the anonymity set of WhatsApp is much larger than that of Signal due to its greater number of users, or how to work out whether this matters. It does not spell out, for example `If you're one of only six people with access to this document it might not be wise to use Signal if none of the other five ever do'. It would also have been prudent to explain that smartphone apps such as Signal and Whatsapp may link the leak to a specific phone number, and that changing your SIM card may not help if this is flagged as suspicious (whether by your social circle or by a network adversary).

The New York Times alternatively suggests sending documents through email, using PGP via the Mailvelope plug-in for Chrome and Firefox. The process for encryption here exhibits similar properties to those that confused most users in the canonical paper of security usability, `Why Johnny Can't Encrypt'~\cite{whitten1999johnny}.  Furthermore, DNS is widely monitored to detect malware and botnets, so DNS access to a commonly-used plug-in for information leakage can flag up a machine as suspicious, especially just after accessing The New York Times. 

Finally, the last option proposed by the New York Times is to use physical post. It recommends using a public mailbox, rather than a post office. It might be a good idea to explain why: the US government scans all snail mail and records sender and destination~\cite{hill2013,nixon_2013_1}, and some letter destinations are more closely monitored than others -- presumably including the New York Times~\cite{nixon_2013_2}. Other risks associated with sending documents through the mail might also be discussed, including new methods for non-invasive content extraction~\cite{redo2016terahertz}.  Lastly, printers themselves represent a vulnerability due identifying watermarks left on documents~\cite{realitywinner} as well as difficult to clear on-device storage. 

Another option mentioned fairly regularly on news organizations is SecureDrop, an application designed specifically for leaking documents. We performed a cognitive walkthrough of this application and identified several usability issues. We believe it would be challenging for a novice user to install the Tor browser, navigate to onion links, and securely store the passphrase that SecureDrop requires them to remember. In fact, using Tor itself can severely reduce a leaker's anonymity set against any adversary that can monitor their network activity.


\section{Whistleblowing Game}

This review taught us that there is no formal model of whistleblowing. In the security literature, well-established ``games'' help shape the mental models of researchers and system designers; all computer science students become familiar with Alice, Bob and Eve playing games with encryption and authentication, and with the Byzantine Generals' Problem for consensus algorithms. 

Perhaps the closest to a whistleblowing game is Simmons' model for covert communications~\cite{Simmons1984}, where Alice and Bob are in jail, and wish to plan an escape; all their communications are monitored by the warden Willie, who will put them in solitary confinement if he can find any evidence of covert communications. The object of the game for Alice and Bob is to communicate in a way that leaves no evidence of its existence, while for Willie it's to prevent this~\cite{Anderson2008}. 

In this section, we propose a whistleblowing game that models a more realistic range of actors.

\subsection{The Actors}
The whistleblowing scenario is more elaborate than the covert communications game:
\begin{enumerate}
    \item \textbf{Alice} is the whistleblower;
    \item \textbf{Duncan} is the reporter;
    \item \textbf{Max} is the boss whose wrongdoing Alice wishes to expose;
    \item \textbf{Tom} is an ally of Max;
    \item \textbf{Harry} is an ally of Alice.
\end{enumerate}

Our final stakeholder is ``the World'', a final arbiter to which the documents may be released. Let us look at the goals for each of these entities:

\begin{itemize}
\item Alice wants to broadcast information about Max to the World;
\item Duncan wants to know that Alice is genuine and, if so, broadcast information to the World;
\item Max wants to stop Duncan from broadcasting this information;
\item Alice wants to plausibly deny her involvement for as long as possible;
\item Max wants to know who Alice is;
\item Tom is an intimidator who can support Max with hard power;
\item Harry is a regulator with hard power who wants to support Alice.
\end{itemize}

\begin{figure}
  \centering
  \includegraphics[width=0.75\textwidth]{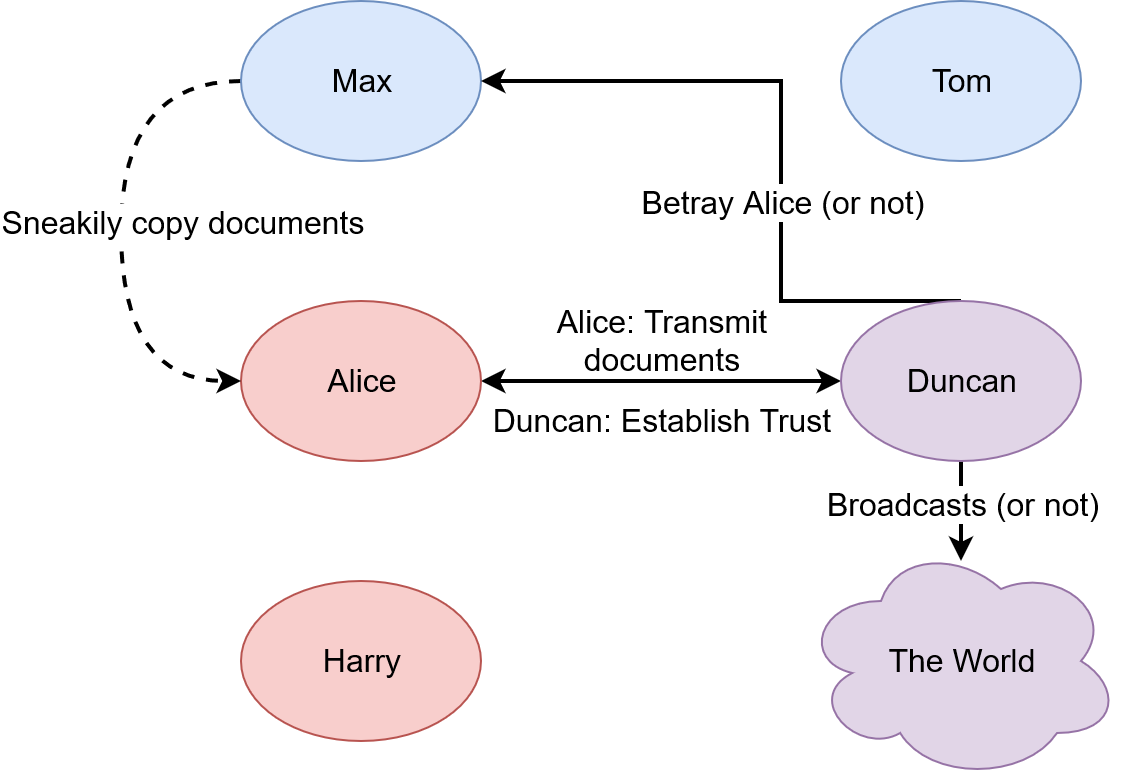}
  \caption{Information flow in the whistleblowing sequence. Entities in red are trying to blow the whistle, entities in blue are their adversaries while the purple entities are ambivalent.}
  \label{fig:wbcomms}
\end{figure}

We suggest that a model of this size may be large enough to reflect real-world tensions while being small enough to be tractable. It is clear what the flow of information in this game looks like, and Figure~\ref{fig:wbcomms} illustrates it. 

The more complex interaction is that of power. Let us see what the flows of power might look like.

\section{Flows of power}

Some insights become obvious once we start thinking of power imbalances. A successful leak is only possible if Duncan, the reporter, can stand up to the miscreant Max; if Duncan can be coerced the leak will not get out to the World, and Alice's cover may be blown too -- if Duncan knows who she is and can be forced to betray her. A second factor is the relative power of Tom and Harry. In a state governed by the rule of law, we would expect that Harry, the regulator, would be much more powerful than Tom, Max's thug, at least in the long run; while in a corrupt state the dictator's henchmen may be able to treat Harry with contempt. So, in a rule-of-law state, the leak may have some intrinsic soft or hard power on its side. If the story is seen as ``just'' or its disclosure ``legal'' then it could be harder for Max to coerce either Alice or Duncan. However, even in wealthy developed countries, an injustice may persist for many years before it catches the public's attention. So any decision by Max as to whether to set his thug Tom on Duncan, or on Alice directly, is likely to involve a calculation of likely consequences.

Let's look at this more closely with a game-theoretic model.

\begin{figure}[h]
    \centering
    \includegraphics[width=\textwidth]{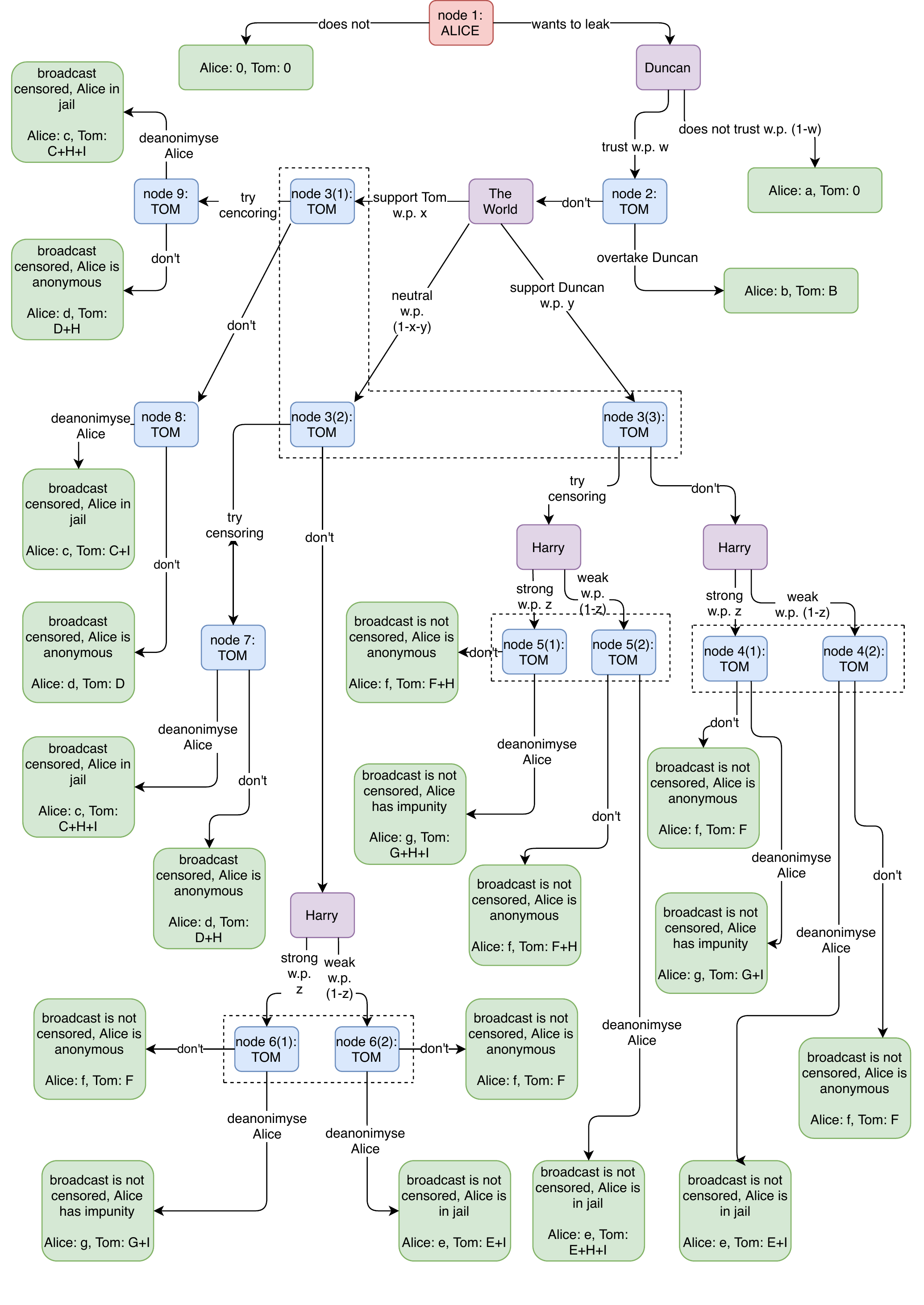}
    \caption{Sequential-move two-player whistleblowing game}
    \label{fig:game}
\end{figure}

Figure \ref{fig:game} presents an extensive form of a stylised whistleblowing game. The game has two players -- Alice and Tom -- and three agents whose strategies are known in advance -- Duncan, the World and Harry. In this setup, Tom is a combination of Tom and Max from the previous section.
	
Alice moves first and has to decide whether she wants to leak the information she has. If she does not, the game ends with both players getting a payoff of 0. If she decides to leak, Alice goes to Duncan, who decides whether or not to trust her. Duncan trusts Alice with a probability $w$ known in advance to both Alice and Tom. If he does not trust Alice, the game ends, again with corresponding payoffs. If he does trust Alice, he broadcasts the information.

What happens next depends on how the World behaves. There are three possibilities: (1) the world supports Tom with probability $x$, or (2) it supports Duncan with probability $y$, or (3) it is neutral with probability $1-x-y$. Tom does not know in advance which course the World will take. We consider these three cases in turn.

(1) The World supports Tom. In this case, the broadcast is in effect censored, regardless of whether Tom spends resources to attempt censoring or not. Subsequently, Tom needs to decide whether to spend resources to de-anonymise Alice. If he does, Alice ends up in jail, and otherwise she remains anonymous. Overall, there are four possible outcomes here, as shown in the figure.

(2) The world supports Duncan. If the world supports Duncan, then the broadcast is not censored, regardless of whether Tom spends resources to attempt censorship. Then Harry comes into the picture and may protect Alice. With probability $z$, Harry's protection is strong , so that Alice ends up with impunity, even if Tom tries to de-anonymise her. With probability $(1-z)$ his protection is weak. In this case, if Tom goes after Alice, she ends up in jail. If Tom doesn't try, she remains free. Overall, there are eight possible outcomes here, each with its respective payoffs for both players.
	
(3) The world is neutral. If the world is neutral, then the broadcast will be censored if Tom spends resources to censor it, and not otherwise. If the broadcast is censored, then Tom has to decide, as in (1), whether he wants to go after Alice. If it is not censored, then Harry comes into the picture, as in (2), and we see six possible outcomes as shown in the figure.
	
This game is solved by backward induction. Payoffs are weighted by relevant probabilities to calculate expected payoffs for both players. Ultimately, Alice decides whether she wants to leak, based on whether the expected payoff of the leak for her is positive or not. The solution turns out to be straightforward if the players are risk-neutral, but it can also be solved assuming both risk-aversion and risk-seeking.

\subsection{Subgame-perfect Equilibrium}

First, let us specify what are the payoffs of outcomes for Tom and Alice, who are the only two players of this game. These are as follows:
\begin{itemize}
    \item Alice does not attempt to leak. both get 0;
    \item Duncan does not trust Alice. Alice gets $a$, Tom gets 0;
    \item If Duncan goes to Max for fact checking, then perhaps Max gets Tom to censor Duncan before the broadcast goes out. Here Alice gets $b$, Tom gets $B$;
    \item Duncan sends out a broadcast but it is quickly censored and Alice ends up in jail. Alice gets $c$, Tom gets $C$;
    \item Broadcast is censored but Alice remains anonymous. Alice gets $d$, Tom gets $D$;
    \item Broadcast is not censored and Alice remains anonymous. Alice gets $e$, Tom gets $E$;
    \item Broadcast is not censored and Alice has impunity. Alice gets $f$, Tom gets $F$;
    \item broadcast is not censored, but Alice ends up in jail. Alice gets $g$, Tom gets $G$.
\end{itemize} 

Throughout the game, Tom has an opportunity to try to censor the broadcast and/or de-anonymise Alice. Tom is uncertain whether these attempts will be successful, since the outcome depends on The World and Harry. Yet Tom knows  how much both attempts will cost him. In particular:
\begin{itemize}
    \item An attempt at censorship costs Tom $H$;
    \item An attempt de-anonymise Alice costs him $I$.
\end{itemize}

Now, to the solution. For simplicity, assume risk-neutrality. Start with the very last decision that Tom has to make -- i.e. whether to de-anonymise Alice. Assume that he will always try to do this if the expected utility is at least as large as that of not doing so. Then, at nodes 7--9, he will try to de-anonymise Alice if $$C + I \geq D$$ Next, nodes 4--6 involve uncertainty over Harry, who can be either support Alice or not with a known probability. Here, Tom will only attempt to de-anonymise Alice if $$zG + (1-z)E + I \geq F.$$ For both types of nodes, determine the preferred action for Tom and label the corresponding expected utilities as $\mathbb{E}(U_n), n \in [4,9].$

Move one step above, to node 3, where Tom has to decide whether he wants to try to censor the broadcast. Again, he will attempt to do so if the expected utility of censorship is at least as large as that of not doing so -- i.e. $$x \mathbb{E}(U_9) + (1-x-y) \mathbb{E}(U_7) + z \mathbb{E}(U_5) \geq x \mathbb{E}(U_8) + (1-x-y) \mathbb{E}(U_6) + z \mathbb{E}(U_4).$$ Determine the preferred action for Tom, and record the expected value of this action as $\mathbb{E}(U_3)$. 

Now move one step above to node 2. Tom will block Duncan if $$B \geq \mathbb{E}(U_3).$$ Determine Tom's preferred action and label it as $\mathbb{E}(U_2)$.

Finally, at node 1, Alice will compare her expected payoff of trying to leak the information with the payoff of not attempting to leak, which we normalise to 0. She expects that with probability $1-w$ Duncan will not trust her, in which case she will receive a payoff of $a$; with probability $w$ he will trust her, in which case her payoff is determined by the actions of Tom as described above. She will only try to leak if her expected payoff of leaking $$(1-w)a + w*\mathbb{E}[\mbox{as determined by Tom's actions}]$$ is positive.

Notice that this model can easily accommodate a case where Duncan goes directly to Harry without first trying to broadcast with the support of the World. Assuming that the broadcast will not be censored, if Harry is involved and the World is not, this requires one to set the probability that the World will support Duncan $y = 1$. Doing this essentially removes nodes 3(1) and 3(2) from the picture as well as all subsequent nodes. A slightly more complex situation is where Alice goes directly to Harry, ignoring both Duncan and the World. This case requires a slight modification in the model in the following way. First, set the probability that Duncan trusts Alice as $w = 1$. Next, Tom's payoff when attempting to overwhelm Duncan should be set to $-\infty$. Finally, as in the previous modification, we set $y = 1$. This will essentially remove Duncan and the World from the setup entirely, and eliminate node (2) from the picture as well as nodes 3(1), 3(2) and all subsequent nodes. The solution method for these two modified cases remain exactly the same as above -- start from the final nodes and go up the tree, at each node determining the preferred action for Alice/Tom. 

\subsection{Revisiting Case Studies}
Let's look at the case studies from Section~\ref{sec:case} again through the lens of this model. In the case of Snowden, we see that the World acted more or less in a neutral fashion. (People in government mostly saw Snowden as a traitor while people in the tech industry mostly saw him as a whistleblower.)  Snowden's adversary had the power to de-anonymize him, but his Duncan had impunity thanks to the US constitution and his Harry, the Russian government, was politically and militarily strong enough to provide Snowden with personal impunity in the form of asylum. We therefore end up with payoffs of $g$ and $G_I$ for Snowden and Tom respectively as shown in Figure~\ref{fig:game}.

We see the World acting in favour of Alice in the Weinstein case, after the New York Times article triggered a shift in public opinion which changed sex abuse by powerful men from being normal to being unacceptable. Then regardless of the presence of a Tom, we would expect the information not to be censored. However, before the World changed its opinion, Tom (in the form of the rich abuser's lawyers) could censor the information with threats of actions for defamation and bills for legal costs. This is exactly what we saw happen in this case.

\section{Recommendations and Future Work}
This analysis suggests that if we wish to help whistleblowers, there are three ways in which technology might be able to help:

\begin{enumerate}
    \item Reducing time-to-publish to zero;
    \item Increasing the cost of de-anonymising Alice;
    \item Facilitating trust establishment between Alice and Duncan.
\end{enumerate}

Variants of reducing the time-to-publish include reducing Duncan's cost-to-publish to zero, increasing his cost-to-betray, and eliminating Duncan completely. Which version you pick may depend on how you characterise ``the World''. When you write a tweet, are you publishing directly to the world or is Twitter now your Duncan? But then there's a further complicating factor, namely the amount of influence you have and thus the likelihood that your tweet will go viral.

Improving Alice's anonymity has been the usual technical approach. As our game shows, increasing her anonymity results in a higher cost for Tom to interfere and thus increases the chances of a successful leak. Moreover, having a reliable means to leak that is well known and easily available could foster a culture of increased transparency~\cite{transint}. This has the second-order effect of increasing Tom's costs in general.

But designers of media websites must be cognisant of the tiny anonymity sets in which most whistleblowers find themselves. This means that bespoke whistleblowing systems may be of suspect utility. Physical intimidation and confiscation of devices are a real possibility, and the mere existence of a leaking app on a person's device may be damning. After all, anonymity loves company, and it may be more effective anyway to teach potential sources to use existing communications tools with more effective operational security. If a newspaper wants sources to use special software to leak, it should be embedded in every single copy of the newspaper's app.

Operationally, the most under-researched avenue of the three mentioned above is the establishment of trust between Duncan and Alice. Not all leaks are bona fide; in a world of increasingly authoritarian governments, news editors must beware of bogus leakers, set up to discredit them. How is a reporter to tell Alice from Malice? What sort of technology might help Duncan establish Alice's good faith without blowing her cover? In the traditional world, trust could be established through mutual friends, or by showing organisational ID, or by disclosing paper documents that would be hard to forge. Electronic equivalents exist but often leave Duncan with digital evidence that might later be seized. Careful use of existing resources can mitigate these risks. For example, Alice might send a photo of her military ID to Duncan by Signal and set the message to disappear after five minutes; that's a decent online equivalent of meeting him in a pub and showing him the real thing. However we see no discussion of such issues on any of the web pages set up to encourage confidential sources to leak to newspapers.

Finally, every introductory lecture on the Prisoners' Dilemma points out that if you don't like the game you find yourself playing, you should try to change it into a different one. Make the dilemma a multi-round game, and you can play tit-for-tat. Can we change the game here?

A relevant case of whistleblowing in the computer industry is the disclosure of software vulnerabilities. Until about 2004--5 this was a Prisoners' Dilemma. If a researcher took a vulnerability to a vendor they might just be threatened with an expensive lawsuit and end up having to promise to keep quiet. The bug meanwhile remained unfixed. So if you wanted to get bugs fixed, you had to take the whistleblowing route and just publish bugs anonymously on bugtraq. That was a lose-lose outcome for everyone, as you didn't get the glory, and the vendor then had to scramble to fix its product, which could get hacked meanwhile. The solution was to change the game to responsible disclosure, whereby the discloser gives the vendor a period of time to patch the vulnerability. Research in security economics suggested this to be the best compromise~\cite{Anderson2008}.

Our own experience of reporting vulnerabilities is that responsible disclosure to non-tech companies can still result in legal threats, and so our standard procedure when we discover a vulnerability in a payment system is no longer to report it to the banks directly but rather to the banking regulators (the Fed, the European Central Bank and the UK Financial Conduct Authority). This was done for example with the vulnerabilities described in~\cite{MDA+2010} and~\cite{BCM+2014}. It removes the threat of legal action against the security researcher, pushes the industry to fix the problem, and with luck embeds compliance tests to ensure that it stays fixed.

If Alice can go directly to Harry, then Duncan becomes redundant. And where Harry's incentives are strong, we often find workable mechanisms. For example, many countries' tax authorities have mechanisms for company staff to blow the whistle on tax avoidance and in many cases to claim quite substantial rewards. However not all industries are regulated well or at all, and even regulated industries have abuses in which their regulators take no interest. So it might be of interest to study which of the world's thousands of regulatory bodies have serviceable whistleblowing mechanisms; this might give useful insights into whether they are serious, or whether they see their real role as protecting the industry they're tasked to supervise. In the context of healthcare, to whom should a concerned medic turn? March 2020 has seen large numbers of medics writing to the press about issues such as the availability of personal protective equipment because of lack of trust in the usual chain of command.

In short, whistleblowing is not just a problem of cryptographic protocol design, or even of usability engineering. It's a complex and fascinating problem in security economics which deserves study in its real-world context.

\bibliographystyle{plain}
\bibliography{sample}
\end{document}